\documentclass[prb,reprint]{revtex4-1}
\usepackage{amsmath,graphicx,amssymb,subfigure,textcomp}

\begin{document}

\title{Mechanism of acoustic emissions from booming sand dunes}

\author{Zhen-Ting Wang}
\email{wangzht@lzu.edu.cn}
\affiliation{Key Laboratory of Desert and Desertification, Cold and Arid Regions Environmental and Engineering Research Institute, Chinese Academy of Sciences, Lanzhou 730000, PRC}

\begin{abstract}
The classical elastic mechanics shows that the fundamental frequency of a sand grain chain is similar to the typical frequency of acoustic emission generated by the booming dunes. The ``song of dunes'' is therefore considered to originate from the resonance of grain chains occurring within a solid layer only several centimeters thick.

\end{abstract}

\maketitle
%


Desert travelers have heard the sounds generated by the booming dunes since Marco Polo's time \cite{Nishiyama82,Nori97,Sholtz97,Patitsas08,Hunt10}. If a person slides down the lee slope of a dune composed of well-polished and well-rounded sand grains, the acoustic emission with the typical frequency of $50-300 Hz$ will be found \cite{Sholtz97}. The sound may continue for several minutes, even after the visible surface avalanche has ceased \cite{Hunt10,Vriend07}. As pointed out by Nori et al. about ten years ago \cite{Nori97}, the greatest attraction of this phenomenon is that it remains an unsolved puzzle. It is well known that pressure fluctuations impinging on the eardrum produce the sensation of sound. Therefore, exploring the grain movements which cause pressure fluctuations in the surrounding air is the key to the origin of the acoustic emissions. The successive grain collisions is frequently regarded as the cause of the sound because the mechanical impulse created by a collision travels as a solitary wave in air \cite{Poynting1922,Bagnold54,Bagnold66,Douady06}. The fundamental period or the inverse of sound frequency is simply the time needed between successive grain collisions when the shear stress remains constant \cite{Poynting1922}. Bagnold advanced the frequency expression in terms of the average grain size and the acceleration of gravity \cite{Bagnold66}, although the chaotic collisions of grains can not emit a sound with a well-defined frequency. Douady et al. suggested that grain movements in the shear layer are synchronized \cite{Douady06,Douady10}. Alternatively, Andreotti et al. proposed that avalanches excite elastic waves at the surface of the dune and the wave-particle mode locking should be responsible for acoustic emissions \cite{Andreotti04,Andreotti08,Andreotti09}. Both of them accepted that the sound frequency is controlled by the shear rate inside the avalanche. However, it seems that the sound frequency as a function of the grain size is not accurate \cite{Vriend08}. In a recent explanation of Hunt and Vriend \cite{Hunt10}, the sand dune was took as a waveguide in which the waves provided by the avalanching and shearing of the surface layer could be reflected at the atmospheric boundary and the substrate half-space. Note the fact that many booming dunes have considerably different types and internal structures, this waveguide model is also incomplete and unsatisfactory.

\begin{figure}[htb]
\centering
\includegraphics[width=0.45\textwidth]{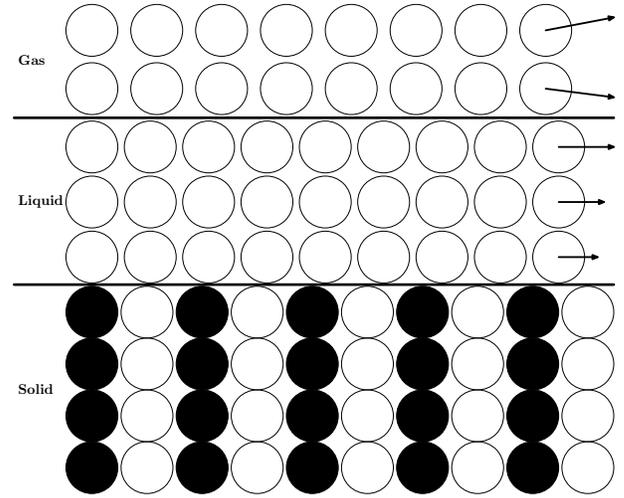}
\caption{three regions in the surface flow of granular material, black: force chains.}\label{surfaceflow}
\end{figure}

In the surface flow of granular media, there exists three distinctive flow regions \cite{Forterre08}: an upper gas region where the grains bounce in all direction, a median liquid region where a dense layer flows, and a lower solid region where grains do not move or creep very slowly. There are various sounds associated with a time-varying air or water flow field around us. Thus, the acoustic sound source is often assumed to locate in the gas and liquid regions in many previous studies, e.g. \cite{Bagnold66,Douady06,Hunt10}. The numerical simulation by directly using equations of fluid mechanics can actually lead to the acoustic frequency spectra which match the experimental data of booming sand \cite{Patitsas03}. But, the later obtained constitutive law for dense granular flow is different from that of water-flow greatly \cite{Jop06}. In the solid region, grains behave like a solid. A vibrating structure can radiate sound waves \cite{Rossing07}. Does the sound output correlate with the solid region?

The forces in the solidified granular media are known to be carried on grain chains (namely ``force chains'') \cite{Cates99}. The well-polished and well-rounded characteristics of booming sand imply us to assume that all force chains are approximatively straight near the bottom of liquid region, as shown in fig. \ref{surfaceflow}. If the moving grains in the liquid region input the energy into such a grain chain in the solid region, one part of energy will transmit along the chain's axis and dissipate eventually, the other part will induce the transverse vibration of the grain chain. Assume that the spherical sand grain is made of perfectly elastic material, the vibration of the chain can be described by the classical elastic theory. In the simplest circumstance of free vibration, the grain chain can be treated as a cantilever beam. Here we numerically solved the three dimensional elastodynamic equation, which can be found in any textbook of elastic mechanics, by using the finite element method and block lanczos method. The contact area between two grains is set to be $s=\pi(0.1d)^2$ where $d$ is the grain diameter.

\begin{figure}[htb]
\centering
\includegraphics[width=0.45\textwidth]{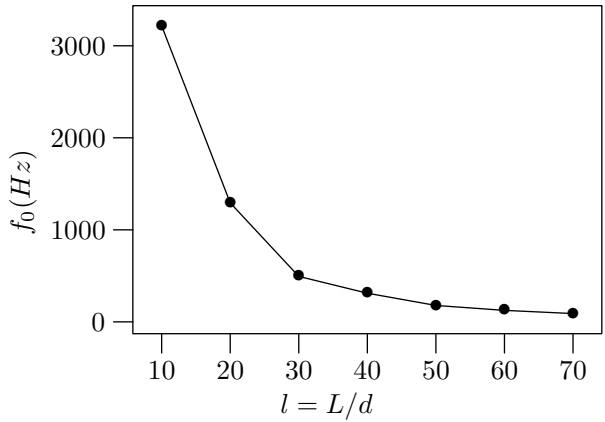}
\caption{variation of the fundamental frequency with the length of grain chain, grain diameter $d=300\mu m$, density $\rho=2.2\times 10^3kg/m^3$, Young's modulus $E=72GPa$, Possion ratio $\nu=0.17$. }\label{f0}
\end{figure}

Fig. \ref{f0} gives the variation of the fundamental frequency $f_0$ with the length $L$ of grain chain. Similar to the frequency of the even mass cantilever beam, $f_0$ rapidly decreases with the increase of $L$. It is surprising that the observed booming frequencies are completely included in the range of $f_0$ when $L=30-100d$. For the booming sand with the diameter of $d=300\mu m$, the needed length of grain chain is only $L\approx1-3cm$. This result is supported by the field measurement of Andreotti et al. which shows that the vibration in the dune is strongly reduced at $6cm$ below the surface \cite{Andreotti04,Andreotti08}. The vibration of grain chain is a building block for the acoustic emission phenomenon. Bagnold proposed that there is an analogy between the sand avalanche and the case of a fingernail being drawn across the corrugations of a book cover \cite{Bagnold54}. Different from his explanation, we think that the frequency of the note emitted is corresponding to the fundamental frequency of the corrugation (or grain chains in the solid region) and the role of finger (or the liquid region) is nothing but providing an instantaneous exciter.

\begin{figure}[htb]
\centering
\includegraphics[width=0.45\textwidth]{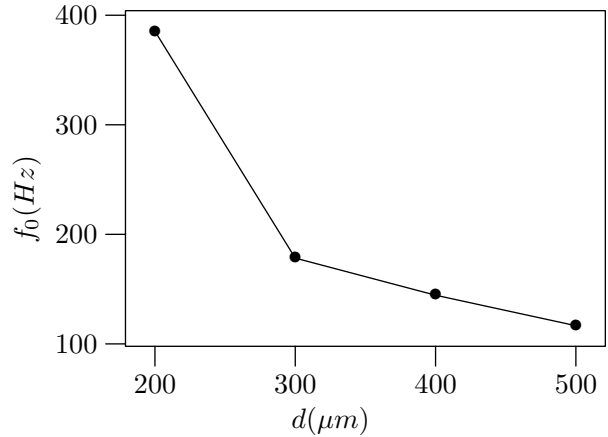}
\caption{variation of the fundamental frequency with the diameter of grain chain, length of grain chain $l=50d$, density $\rho=2.2\times 10^3kg/m^3$, Young's modulus $E=72GPa$, Possion ratio $\nu=0.17$. }\label{df}
\end{figure}

Some researchers argued that the booming frequency should be proportional to $\sqrt{g/d}$ where $g$ is the acceleration of gravity \cite{Bagnold66,Douady06,Douady10}. An analogous $f_0-d$ relation was also obtained in this study, see fig \ref{df}. Grain size plays an essential role in the aeolian process because the aerodynamic force is sensitive to it. But, the narrow size scope of booming sand and the slow variation of $f_0$ weaken the importance of this parameter in the current process.

The resonance of grain chains occurring within a solid layer only several centimeters thick is a novel explanation for acoustic emission of booming sand, although the experimental evidence is insufficient. To our knowledge, there are three possible physical mechanisms responsible for the formation of grain chain. First, Douady et al. attributed their ``spatial coherence'' in the granular flow to friction \cite{Douady06,Douady10}. Second, the cluster formation in freely falling granular streams is due to nanoNewton cohesive forces \cite{Royer09}. Third, the length of grain chain was observed to be $1.3cm$ about while slowly pouring booming sand \cite{Lewis36}. The electrical charge, which was verified to exist on these grain chain by an electroscope \cite{Sholtz97}, was thought to make on grain adhere to another. The answer provided here is simple and crude. We expect it can stimulate follow-up works to completely reveal the mystery of this interesting natural phenomenon.

\begin{acknowledgments}
This research was supported by NSFC project (No. 11274002).
\end{acknowledgments}

\end{document}